# Short and soft: multi-domain organization, tunable dynamics and jamming in suspensions of grafted colloidal cylinders with small aspect ratio

*Daniele Parisi[1,2], Yingbo Ruan[3,4], Guy Ochbaum[5], Kevin S. Silmore[6], Lucas L. Cullari[5], Chen-Yang Liu[3,4], Ronit Bitton[5], Oren Regev[5], James W. Swan[6], Benoit Loppinet[1] and Dimitris Vlassopoulos[1,2]*

1 Institute of Electronic Structure & Laser, FORTH, Heraklion 71110, Crete, Greece

2 Department of Materials Science & Technology, University of Crete, Heraklion 71003, Crete, Greece

3 Beijing National Laboratory for Molecular Sciences, CAS Key Laboratory of Engineering Plastics, Institute of Chemistry, The Chinese Academy of Sciences, Beijing 100190, China

4 University of Chinese Academy of Sciences, Beijing 100049, China

5 Department of Chemical Engineering and the Ilze Katz Institute for Nanoscale Science & Technology, Ben-Gurion University of the Negev, Beer-Sheva 84105, Israel

6 Department of Chemical Engineering, Massachusetts Institute of Technology, Cambridge, MA, 02139, USA

**Abstract**

The yet virtually unexplored class of soft colloidal rods with small aspect ratio is investigated and shown to exhibit a very rich phase and dynamic behavior, spanning from liquid to nearly melt state. Instead of nematic order, these short and soft nanocylinders alter their organization with increasing concentration from isotropic liquid with random orientation to one with preferred local orientation and eventually a multi-domain arrangement with local orientational order. The latter gives rise to a kinetically suppressed state akin to structural glass with detectable terminal relaxation, which, on increasing concentration reveals features of hexagonally packed order as in ordered block copolymers. The respective dynamic response comprises four regimes, all above the overlapping concentration of 0.02 g/ml: I) from 0.03 to 0.1 g/mol the system undergoes a liquid-to-solid like transition with a structural relaxation time that grows by four orders of magnitude. II) from 0.1 to 0.2 g/ml a dramatic slowing-down is observed and is accompanied by an evolution from isotropic to multi-domain structure. III) between 0.2 and 0.6 g/mol the suspensions exhibit signatures of shell interpenetration and jamming, with the colloidal plateau modulus depending linearly on concentration. IV) at 0.74 g/ml in the densely jammed state, the viscoelastic signature of hexagonally packed cylinders from microphase-separated block copolymers is detected. These properties set short and soft nanocylinders apart from long colloidal rods (with large aspect ratio) and provide insights for fundamentally understanding the physics in this intermediate soft colloidal regime, as well as and for tailoring the flow properties of non-spherical soft colloids.

## INTRODUCTION

Whereas spherical colloids undergo a transition from liquid to crystalline phase on increasing volume (or weight) fraction (with possible intervention of a wide glass region, depending on polydispersity and softness[1] which is often materialized via particle grafting with a polymeric coat[2,3]), long rods exhibit liquid crystalline mesophases before packing into fully hexagonal order likely.[4] Interestingly, recent experimental evidence with fd-virus[5–7] has demonstrated that on increasing their volume fraction, the rods form first nematic domains, which subsequently freeze into a glassy state characterized by two length scales (rod and domain) and then give rise to re-entrant liquid crystalline order, with a respective variation of dynamic properties. With the help of simulations it has been shown that further increase of the volume fraction transforms the glass into extended nematic, small smectic and/or phase separated regimes, before eventual equilibration into crystalline order.[8] Softness typically smears out crystal ordering transitions as shown repeatedly with spherical colloids such as star polymers (although crystalline order may develop after long waiting times).[9,10] In such a case, at high volume fractions, overlapping of the polymeric coats plays a key role in structure and dynamics. The extent of mesophases and glassy regime, as well as their interplay with phase separation, strongly depend on the aspect ratio of the rods, as demonstrated with simulations[8,11,12]. The latter have revealed the complex multi-domain structure of short hard rods: at high enough volume fractions they form multi-layered crystalline clusters, whereas at higher concentrations small crystallites arrange into long-lived structures with local orientational order albeit kinetically arrested. At even higher concentrations, glass transition of the colloidal clusters takes place. Two distinct features at small aspect ratios (but not too small, say in the range 5-10) are (i) the weaker mesophase transition, characterized by shorter range interactions (typically within the clusters) and (ii) a tendency for kinetic arrest, which masks phase separation and crystallization. Introducing softness into these short cylinders (by means of grafting polymer chains) provides the means to alter their spatiotemporal response by coupling their multi-domain structure with the polymeric features of the grafts, notably their interpenetration. This enhances our ability to tailor the flow properties of soft non-spherical colloids, with substantial implications in processing applications.[13–15] There are only a few works exploring the combined roles of softness and anisotropy, mainly with microgels in two and three dimensions, and are focused on their structural features under electric field or confinement.[16–20] Grafting long colloidal rods has been explored as a means to obtain soft anisotropic colloids with very long aspect ratios, whose structure and rheology have been explored focusing on the interplay of nematic order and thermoreversible gelation due to the response of the grafted chains.[21]

The polydispersity has a significant effect on the phase behavior of rod-like systems. Bates and Frenkel[22] reported that infinitely long rod-like colloids exhibit an unchanged phase behavior if the dispersity in length is less than 8%. For polydispersity in the range 8-18%, the smectic phase was found to become increasingly destabilized with respect to the nematic and columnar phase at low and high concentrations, respectively. Above 18% polydispersity, only nematic and columnar phases could be observed. Lekkerkerker and Vroege[23] showed that in bidisperse infinitely thin rods the isotropic-nematic coexistence region broadens on increasing polydispersity. This was also observed experimentally by Buining *et al.*[24] in suspensions of boehmite rods. Speranza and Sollich[25] also investigated the phase behavior of hard thin rods with bimodal length distributions. They found a three-phase isotropic-nematic-nematic coexistence region when the asymmetry in length between long and short rods was sufficiently large. Same results were also observed by Wensink and Vroege.[26] Less is known about the polydispersity effect on the phase behavior in cylinders with small aspect ratio. The reason lies on the fact that small aspect ratios (below 8) destabilize mesophases, even for monodisperse rods, as shown by Maeda and Maeda experimentally[27] and Lekkerkerker and Dijkstra through simulations.[23,28] For very short rods (or cut spheres) Blaak *et al.*[29] reported that polydispersity favors cubatic phase instead. Such a phase was also predicted for charged rods in the presence of inter-rod linkers.[30]

The effects of polydispersity and aspect ratio on the phase diagram of rod-like colloidal systems are reflected in their dynamics as well. Dhont *et al.*[7] showed that fd-virus suspensions at low ionic strength exhibit a nematic-to-glass transition, are characterized by two length scales (rod and domain) and exhibit dynamic features of a yield-stress fluid. Wierenga and Philipse[31] and Solomon and Boger[32] investigated the dependence of the shear viscosity on the Debye length and volume fraction of charged boehmite rods (with aspect ratio of about 22.5). The dynamics of hairy-rod polyesters with molar mass polydispersity of 1.7 was found to be dominated by clusters with only local order as revealed by birefringence measurements[33,34]. Various levels of organization and kinetic frustration in colloidal rods have been reported in the literature, theoretically and experimentally[35–38]

Taking advantage of the properties of microphase separated block copolymers has been shown to be a promising route for obtaining colloids with systematically varied size, shape and softness, which exhibit interesting tunable rheology[39,40], however the dynamic properties of the so-obtained non-spherical grafted colloids have not been explored. In this work we exploit this avenue. By using well-characterized grafted cylinders with aspect ratio of about 6, we employ of suite of experimental and simulation techniques to investigate their structure and dynamics as functions of concentration from liquid to almost molten state. We show the respective transition from isotropic liquid to small domains with local orientational order, to a soft glass with local orientation and detectable terminal relaxation (cage escape), and eventually a hexagonally-like packed arrangement. We identify the distinct features of such short soft cylinders. These include (i) multi-domain dynamic response with distinct scaling, (ii) decoupled short-time polymeric and long-time colloidal response at high weight fractions, and (iii) different packing arrangement reflected in strong dependence of viscoelastic properties on weight fraction, bearing signatures of interpenetration and jamming.

## MATERIALS AND METHODS

**Materials**

We used nanorods consisting of crosslinked poly(3-(triethoxysilyl)propylmethacrylate) (PTEPM) core and linear polystyrene (PS) grafted chains, made of PTEPM-PS asymmetric block copolymers, microphase separated into a cylindrical phase with the core subsequently crosslinked.[41] Such an innovative synthesis process is called Assembly-Cross-link- Dispersion (ACD), which allows the fabrication of precisely controlled nanoparticles. These particles may have varied shapes but identical PS shells, varied core sizes but the same PS shell, or fixed shapes but varied PS shells.[41] A simplified cartoon of the original copolymer and the resulting cylindrical structure is shown in Figure 1. The PS segments of the diblock constitute the shell of the cylinder, whereas the crosslinked PTEPM segments represent the core. Further details on the synthesis are reported in Ref. 42.

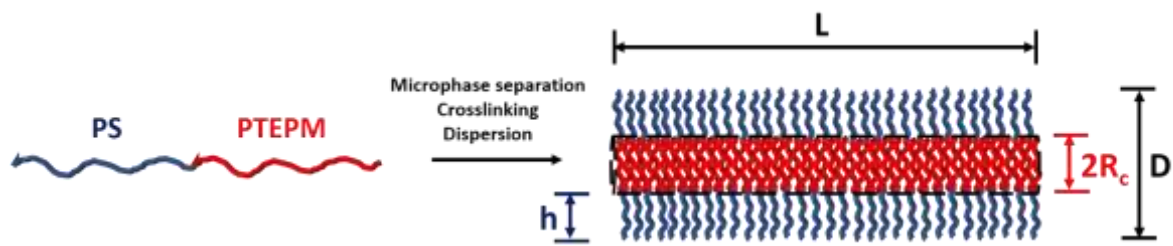

Figure 1. Schematic illustration of a grafted nanocylinder (right). The PS-PTEPM copolymer on the left-hand side constitutes the building block component of the nanoparticle.

The nanorods used in this work are characterized by the following properties. The number-average molar mass of PS is $M_{PS,arm}$ =111 kg/mol, their grafting density σ=0.102 chains/nm² and the respective weight fraction 79%. The PTEPM core radius ($R_c$) is 13.5 nm (based on TEM), the calculated PS shell (h) is 56 nm thick (see Supporting Information, SI, section SI1), leading to a total diameter (D) of 139 nm. The length (L) of the cylinder is 770 nm (from TEM) and the resulting polydispersity index PDI is 1.3, yielding an average aspect ratio of 6 with a prevalently stiff conformation (given the highly crosslinked core).[41] The value of PDI was confirmed by analysis of the form factor (see SI, section SI2). The total molar mass of a cylinder was estimated from the total molar mass of the PS chains in the shell $M_{PS,shell}$= σ x L x 2π$R_c$ x $M_{PS,arm}$ and the PS mass fraction in a cylinder (79 wt%), yielding $M_{cylinder}$=$M_{PS,shell}$/0.79= 9.3x10⁵ kg/mol. The overlap concentration can be defined as C*=$M_w$/L²(2$R_c$+2h)$N_A$, with $M_w$ being the weight-average molar mass and $N_A$ the Avogadro number. The main molecular characteristics and sizes are summarized in Table 1 below.

Table 1. Molecular characteristics of the nanocylinders investigated

| Sample | Number of arms | $M_n$ [kg/mol] | PDI [-] | L [nm] | D [nm] | C* [g/ml] | L/D [-] |
|---|---|---|---|---|---|---|---|
| Nanocylinders | 6600 | 9.3x10⁵ | 1.3 | 770 | 139 | 0.02 | 6 |

Suspensions in diethyl phthalate (DEP) were prepared for cryo-TEM, X-rays, light scattering and rheology, whereas toluene (Tol) was used as a solvent for cross-polarized microscopy imaging and light scattering. In addition, Tol offers good contrast with PS whereas it is nearly index-matched with silica and is athermal solvent for PS which represents the main particle constituent.[42–44] DEP, which is good solvent for PS in the investigated temperature,[44] was used in long-time experiments (especially rheological) because it is much less volatile (its boiling point at atmospheric pressure is 295ºC). Hence, the PS shell is swollen in both solvents (see also SI, section SI3). A wide range of concentrations from dilute to the highly concentrated regime was investigated. The effective volume fraction of the nanocylinders can be estimated either as the ratio C/C* or by considering an equivalent hard cylinder with the same total molar mass of the grafted nanocylinder. To this end, the size of such cylinder was found by assuming $M_{cylinder}=M_{PS}$, therefore $R_{hard\ cylinder}=M_{PS}/(\sigma \times L \times 2\pi)=19$ nm. From the knowledge of the mass concentration C and the molecular characteristics, i.e., $M_{cylinder}$, L and $R_{hard\ cylinder}$, the equivalent volume fraction is calculated as: $\phi = C(N_A/M_{cylinder})(\pi R^2_{hard\ cylinder}L)$. The respective values are listed in Table 2 below. Note that the $\phi$ values are the same as those used in simulations (see section on structure and phase behavior below).

Table 2. Estimated volume fractions of the nanocylinders investigated

| Concentration (g/ml) | 0.03 | 0.1 | 0.13 | 0.2 | 0.23 | 0.4 | 0.6 | 0.76 |
|---|---|---|---|---|---|---|---|---|
| C/C* | 1.5 | 5 | 6.5 | 10 | 11.5 | 20 | 30 | 37 |
| $\phi$ | 0.016 | 0.054 | 0.070 | 0.107 | 0.123 | 0.214 | 0.321 | 0.407 |

**Dynamic light scattering (photon correlation spectroscopy)**

Photon Correlation Spectroscopy measurements were performed in two different instruments depending on the suspension concentration. In the very dilute regime (C≤$10^{-3}$ g/ml) with the solutions being completely transparent, an ALV-5000 goniometer/ correlator setup equipped with a Nd:YAG laser (130mW) at λ=532 nm, as the light source, was used. The experimental normalized light scattering intensity autocorrelation function[45] $G(q,t) \equiv \langle I(q,t)I(q)\rangle/|\langle I(q)\rangle|^2$ was measured over a broad time range ($10^{-7}$–$10^3$ s) at different values of the scattering wave vector $q = (4\pi n/\lambda)\sin(\theta/2)$, where $n$ is the refractive index of the medium and $\theta$ the scattering angle (which was varied in the range between 30° and 150°). Under homodyne beating conditions, the desired concentration fluctuations relaxation function (or intermediate scattering function, ISF) is computed from the experimental $G(q,t)$ as $C(q,t) = [(G(q,t)-1)/f^*]^{1/2}$, where $f^* = 0.36$ is an instrumental coherence factor. In the dilute regime $C(q,t)$ is a single decay function and the effective short-time diffusion coefficient is determined from the initial decay rate Γ according to[45,46]

$$D_{Sh} = \frac{\Gamma}{q^2} = \left(\frac{1}{q^2}\right)\lim_{t=0}\left(\frac{d[\ln C(q,t)]}{dt}\right) \qquad (1)$$

Details are provided in the SI, section SI4. At higher concentrations, solutions become turbid and multiple scattering events that take place represent an issue. To minimize this undesired phenomenon we used a 3D Dynamic Light Scattering setup (3DDLS from LS Instruments)[47–50] equipped with a He-Ne laser (22.5mW) at λ=632.8 nm for the two laser beams. The scattering angle was varied between 30° and 135°. The cross-correlation scheme yields the following normalized cross-correlation function:

$$g_c^{(2)}(\tau) = \frac{3\langle I_1\rangle\langle I_2\rangle + \langle I_1^i(0)I_2^{ii}(\tau)\rangle}{\langle I_1^i+I_1^{ii}\rangle\langle I_2^i+I_2^{ii}\rangle} = \frac{3}{4} + \frac{1}{4}[1+\beta^2[C(q,\tau)]] = 1 + \beta_{tot}^2[C(q,\tau)] \qquad (2)$$

where $I_1$ and $I_2$ represent the collected intensities by the first and second detector respectively, the superscripts indicate the first and the second laser beam, $\beta_{tot}^2 = \beta^2 \beta_{OV}^2 \beta_{MS}^2 \beta_T^2$ is the correction factor, with $\beta^2$ being the "coherence factor" which is related to coherence area and depends on the detection optics, $\beta_{OV}^2$ the "overlap factor" that accounts for the fact that the scattering volumes seen by the two detectors are slightly different, $\beta_{MS}^2 = \frac{(g^{(2)}(q,\tau=0)-1)_{conc}}{(g^{(2)}(q,\tau=0)-1)_{dilute}} = \frac{\beta_{conc}^2}{\beta_{dilute}^2}$ the "multiple scattering factor" that is determined as the ratio of the amplitude of the cross-correlation function for the concentrated sample to the amplitude of the correlation function for a dilute sample and $\beta_T^2$ the "technique" factor connected to the "unused erratic" light collected by the detectors ($\beta_T^2 = 0.25$ for the 3DDLS set-up used here). All experiments were carried out at T=25°C, and in both instruments the bath temperature was controlled by means of a heat exchanger coil which was connected to a thermostat. The same experimental apparatus was used to measure light transmission.

**Transmission Electron Microscopy (TEM)**

Room temperature (RT) TEM images were obtained using a (JEOL-1011) TEM instrument operated at an accelerating voltage of 100 kV. The images were recorded by a Gatan 794 CCD camera. For observing the morphology of nanoparticles in dry state, a drop of the suspension was deposited onto a copper grid (Formvar stabilized with carbon support films) and solvent evaporated at room temperature before the TEM analysis was performed. Further details are provided in the SI, section SI5.

**Cryogenic-Transmission Electron Microscopy (cryo-TEM)**

Cryo-TEM images were collected using a FEI Tecnai 12 G2 (equipped with a Gatan 794 CCD camera) operated at 120 kV. The samples were prepared using a controlled environment vitrification system. A drop of the sample was deposited on a TEM grid (Lacey substrate 300 mesh, Ted Pella, Ltd.) at 25 °C. The grid was blotted to remove excess fluid, resulting in the formation of a thin film (20–300 nm) of the solution suspended over the grid's holes. The samples were vitrified by rapid plunging into liquid ethane at its freezing point, and then transferred to liquid nitrogen and to a cryo holder (Gatan model 626) for imaging.

**Small-Angle X-ray Scattering (SAXS)**

SAXS measurements were performed with a GANESHA 300-XL instrument (SAXSLAB). CuK$_\alpha$ radiation was generated by a Genix 3D Cu-source with an integrated monochromator, 3-pinhole collimation and a two-dimensional Pilatus 300K detector. The scattering intensity $I(q)$ was recorded in the interval of $0.003 < q < 3$ Å$^{-1}$ (corresponding to the length range of 3-200 Å) where the scattering vector is defined as $q = 4\pi \sin\theta / \lambda$, with $2\theta$ and $\lambda$ being the scattering angle and wave length, respectively. The measurements were performed under vacuum at room temperature (RT). Solutions at $10^{-4}$, $10^{-2}$, 0.1, 0.13, and 0.2 g/ml were placed in a stainless-steel cell with entrance and exit windows made of mica. The scattering curves were corrected for counting time and sample absorption. No attempt was made to convert the data to an absolute scale. See also SI, section SI6.

**Shear Rheology**

To investigate the viscoelastic properties of the suspensions, rheological experiments in the linear viscoelastic regime were performed using a strain-controlled rheometer ARES (TA Instruments) with a force rebalance transducer 100FRTN1 and a cone-plate geometry (stainless steel 8 mm diameter, cone angle of 0.166 rad). A stress-controlled rheometer MCR 501 (Anton-Paar) equipped with cone-plate geometry (stainless steel 8 mm, cone angle of 0.017 rad) was used for creep experiments (see also SI, section SI7). Temperature control was achieved by means of Peltier elements placed at the lower plates of the rheometers. All experiments were performed at 25 °C. A home-made solvent trap was used in all experiments to seal the sample from the external environment (unwanted air current) and create a closed saturated atmosphere (note that the high boiling point of DEP did not create evaporation concerns anyway). The measurement protocol involved the following tests: i) dynamic strain sweep at 1 rad/s from low to high strain amplitudes (typically from 0.001 to 3 strain units) to detect the linear viscoelastic regime and rejuvenate the system; ii) dynamic time sweep in the linear regime at 1rad/s and 0.005 strain units until steady-state of the dynamic moduli was reached (typical waiting time for concentrated suspensions was of the order of 30 hours); iii) dynamic frequency sweep at 0.005 strain units over a range of frequencies between 0.01 and 100 rad/s, depending on the torque resolution. Subsequent creep experiments were conducted following the abovementioned protocol. Different stresses were applied to linear response. Creep compliance was then converted into dynamic moduli by means of the nonlinear regularization method of Weese[51] in order to extend the low-frequency region. As discussed in Section III.2 below, some high-frequency (above 100 rad/s) measurements were performed at selected high concentrations by means of a home-built piezo-rheometer (see SI, section SI8).[52]

**Simulations**

Brownian dynamics simulations of linear, rigid assemblies of beads were conducted using the HOOMD-blue simulation package[53,54] with periodic boundary conditions in order to model the behaviour of the experimental nanorods featured here. The number of beads comprising the simulated rods (and hence their length) was drawn from a lognormal distribution (a commonly encountered particle size distribution) in order to recreate the experimentally measured PDI of 1.3. Given that the exact interparticle interactions in our experimental systems are not known, we used a soft repulsive potential to test how variations in softness influence the microstructure (see also the SI, section SI9). This potential is infinitely repulsive within a hard core radius and is repulsive between the hard core radius and the graft radius with a "softness" governed by the exponent of a power law. Static structure factors were calculated using a non-uniform fast Fourier transform[55] and were averaged over 251 independent simulation snapshots. Further technical details can be found in the SI (section SI9).

## RESULTS AND DISCUSSION

### Structure and phase behavior

The phase behavior from dilute to concentrated regime is mapped in Figure 2 which depicts the zero-shear viscosity (in DEP), translational diffusion coefficient (in Tol), light transmission (in Tol) and optical appearance through crossed polarizers (in Tol) as a function of particle concentration. At very low concentrations in the dilute regime (up to 0.041 g/ml in Tol), the system is isotropic and characterized by normalized translational diffusion coefficient and zero-shear relative viscosity being identical and practically concentration-independent. As the concentration increases, the diffusion and inverse viscosity slow-down by the same degree. Concomitantly, the transmitted light intensity decreases, suggesting that the solution (in Tol) becomes more turbid, as corroborated by visual observation. Nevertheless, the cylindrical particles do not exhibit long-range ordering as there is no unambiguous evidence of birefringence. At higher concentrations, diffusion and viscosity exhibit a stronger (albeit same) dependence on concentration, whereas a weak birefringence is also observed. The isotropic phase coexists with small domains where the cylinders are locally oriented as sketched in the cartoon (0.049 g/ml in Tol). Further increase in concentration at about 0.1 g/ml in Tol, leads to a non-ergodic state (detected with dynamic light scattering, see Figures S3 and S4 of the SI) where the nanocylinders are kinetically trapped into locally favoured structures.[56] In this situation, the dynamics is dramatically slowed-down. Small aspect ratio and polydispersity in length are known to promote this arrangement.[11,12,57] The dynamic arrest is accompanied by an abrupt drop of the inverse reduced viscosity and an increase of the light transmission value. The latter is rationalized by the fact that, since the scattering intensity is thought to originate primarily from the contrast between the polymeric shell (about 80wt% of the total mass of the particle) and the solvent, its reduction upon increasing concentration should reflect the approach of neighboring particle shells. Note that the largest refractive index difference is between the shell and the core, but since the latter constitutes only 20 wt% of the particle, its effect is rather small. Remarkably, the experimental light transmission values, measured in the arrested state at different positions of the vial, were nearly the same. This conforms to a macroscopically homogenous sample where nanocylinders are locally oriented in domains of random direction. Concentrations above 0.1 g/ml in Tol exhibit a non-monotonic variation in birefringence, which eventually tends to vanish at 0.29 g/ml. Along with the continuous increase of transmission (albeit weaker at the highest concentrations), this suggests that on increasing concentration the system forms a multi-domain structure with a stronger shell-shell engagement between adjacent nanocylinders. In order to further elucidate the structure of the different phases at different concentrationes, complementary SAXS measurements were performed.

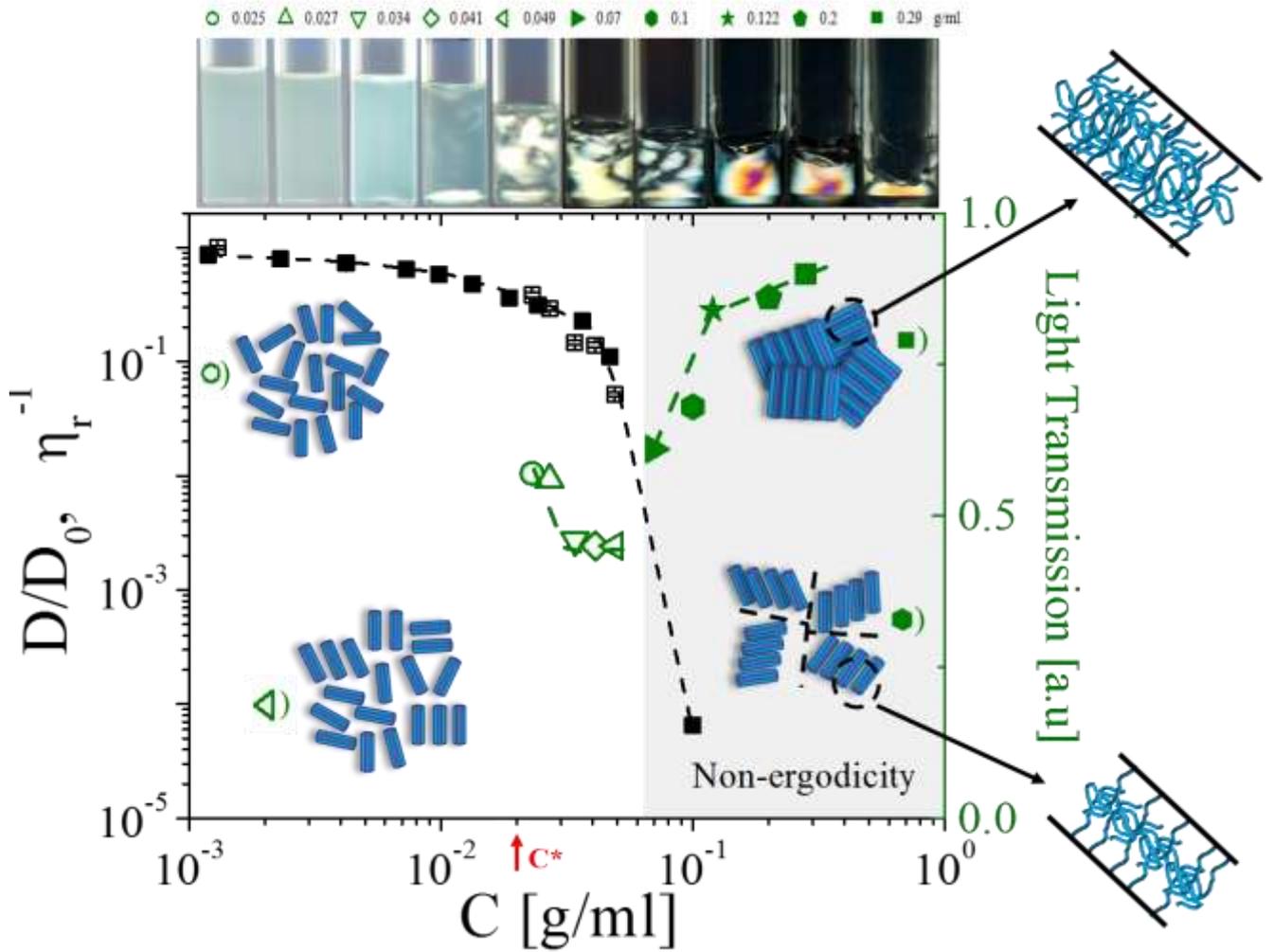

Figure 2. State diagram of nanocylinder suspensions. Top panel: cross-polarized images at different concentrations in toluene. Bottom panel: normalized translational diffusion coefficient (□), inverse relative viscosity (■), and light transmission (right y-axis) with green symbols (corresponding to the top panel images) against concentration. The overlapping concentration in DEP (see Table 1) is indicated by a red arrow (0.02 g/ml). Symbols next to the schematics refer to the corresponding concentration in the cross-polarized images (in toluene). Dashed lines are drawn to guide the eye. The cartoons illustrate the different structural arrangements and the magnifications show the different degrees of grafted layer interpenetration in the jammed regime (see text).

Indeed, the evolution of the SAXS 2D-intensity patterns with the concentration is shown in the top panel of Figure 3A. Overall, these patterns point to macroscopically isotropic solutions, consistent with Figure 2, as well as results from simulations[8] and experiments[58] with short hard cylinders. The intensity curves superimpose in the high-q-regime (0.2-2 nm$^{-1}$) when the intensity is scaled with the particle concentration. However, the data in this regime, which reflect shell dynamics, are not well-resolved and lie between Gaussian ($I \sim q^{-2}$) and self-avoiding ($I \sim q^{-1.69}$) behaviors (Figure 3A). At lower q-values, the intensity curves do not overlap so well anymore, reflecting the effects of particle interactions. The deviation is first evidenced at 0.1 g/ml and becomes more apparent at higher concentrations. Structural peaks appear and their position is affected by the concentration as the value of q at the first peak (q*) slightly increases with concentration. Concerning the latter, assuming that the total scattering intensity can be expressed to a first approach as I(q)=S(q)P(q),[45,59] the extracted pseudo-structure factor S(q) at different concentrations is depicted in the top inset of Figure 3A and reveals up to 4$^{th}$ order peaks at 0.1, 0.13 and 0.2 g/ml. It represents a signature of a long-range positional order in the measured q-range. Such a long-range order with relative peak positions at q*, 2q*, 3q*, and 4q* is reminiscent of cubatic order.[12] Moreover, while polydispersity is known to destabilize crystalline phases, on the contrary it can favor the stabilization of a cubatic phase as demonstrated by Blaak.[57] A similar particle arrangement can be thought in our system, given also the fact that some polydispersity is present. As the concentration increases from 0.13 to 0.2 g/ml, the structural peaks slightly move to higher q-values (the nanocylinders come closer). Whereas for the static structure this does not seem to be an important variation, it has a dramatic impact on the dynamics as particle shells start to interpenetrate and linear chains form an effective polymeric network. Shell interpenetration at high concentrations for the given grafting density and molar mass of the grafted chains is expected in order to minimize local density heterogeneities (see also the SI, section SI10).[60,61] At 0.1 g/ml, the structural peaks are not clearly detectable, indicating that the structure is not fully developed, whereas an isotropic-multi-domain phase coexistence is present instead. We conclude that the transition detected by means of birefringence is supported qualitatively by SAXS.

RT- and cryo-TEM micrographs are shown as insets in Figure 3A (additional TEM images are shown in Figure S5 of the SI). In the dry-state (RT-TEM), a multi-domain organization of the nanocylinders is clearly observed, with a random coordination number within each domain. The latter has implications on the difficulty encountered to experimentally determine a characteristic size of the domains, as well as the absence of detectable relaxation dynamics of such domains discussed below. Due to the drying process, the structure visualized in RT-TEM imaging might be affected. For this reason, to corroborate the aforementioned scenario cryo-TEM images are also shown in Figures 3A and S5. Although the contrast is not as good as with the dry-TEM, due to lower electron density difference (mass-thickness contrast), the presence of a multi-domain arrangement of the soft nanocylinders with no preferential orientation is validated.

The structure factor from simulations of the 0.4 g/ml suspension is shown in Figure 3B as a function of the softness of the repulsive potential between the rods. The simulations use a log-normal length distribution to replicate the experimental polydispersity. Therefore, in the simulations no crystalline order is observed, in agreement with the experimental results. We argue that there is qualitative agreement with experiments (Fig.3A inset). However, when the softness parameter of the potential k approaches ∞ (see SI, section SI9), the potential approaches that of a hard spherocylinder, as considered by Yatsenko and Schweizer.[36] We expect in that limit that we may obtain a stronger "sawtooth-like" correlation. Two additional remarks are in order here. First, the structure factor computed in the simulations (and in ref. 37) is the scattering intensity from all the beads in the rod divided by (the intra-rod) scattering intensity of a single "bead-rod". This may mask or enhance the appearance of oscillations in this site-site structure factor on length scales proportional to the bead size. Second, the pseudo-structure factor S(q) computed from experimental measurements as discussed above, is a different quantity from that calculated in simulations (where the site-site structure factor is based on a uniform scattering cross section along the length of a rod). The effect of softness on the orientational correlation for nanocylinders at 0.4 g/ml is shown in Figure 3C. The orientational correlation function $S(r)$ based on center of mass of rods is defined as

$$S(r) = \frac{3}{2}\langle(\hat{n}(0)\cdot\hat{n}(r))^2 - \frac{1}{3}\rangle \qquad (3)$$

where $\hat{n}(r)$ is the orientation vector of the cylinder. The reported values of $S(r)$ are orders of magnitude lower than the value expected for pure nematic phase, 1 (a value of 0 corresponds to isotropic system).[62] Increasing stiffness appears to yield an increased ordering at moderate distances, but the effect is very weak. To a first approximation, we infer that such nanocylinder systems display a macroscopically isotropic structure. However, it should be noted again that simulated and experimental orientation order parameters are different. In the simulations, we are examining how correlated the rods are with respect to some inter-rod spacing and find that the correlation, however big, dies off fast in the simulations. Figure 3D shows snapshots of simulation boxes at various concentrations. Interestingly, between 0.03 and 0.1 g/ml it is possible to observe a transition towards the formation of a highly isotropic solid-like structure. Such a transition has been also observed by means of rheological experiments, as it will be shown next. In addition, at higher concentration (0.4 g/ml), the nanocylinders are severely crowded and homogeneously distributed throughout the simulation box, in harmony with the results from X-ray measurements. However, no distinct signatures of order within the domains, where cylinders are locally organized, have been observed with simulations. The reason for this may be the repulsive, coarse-grained potential used, which despite mimicking softness, does not take into account the presence of grafted chains or any possible weak attractive interaction, hence the possibility of chain interpenetration which might affect particle arrangement and dynamics.[63]

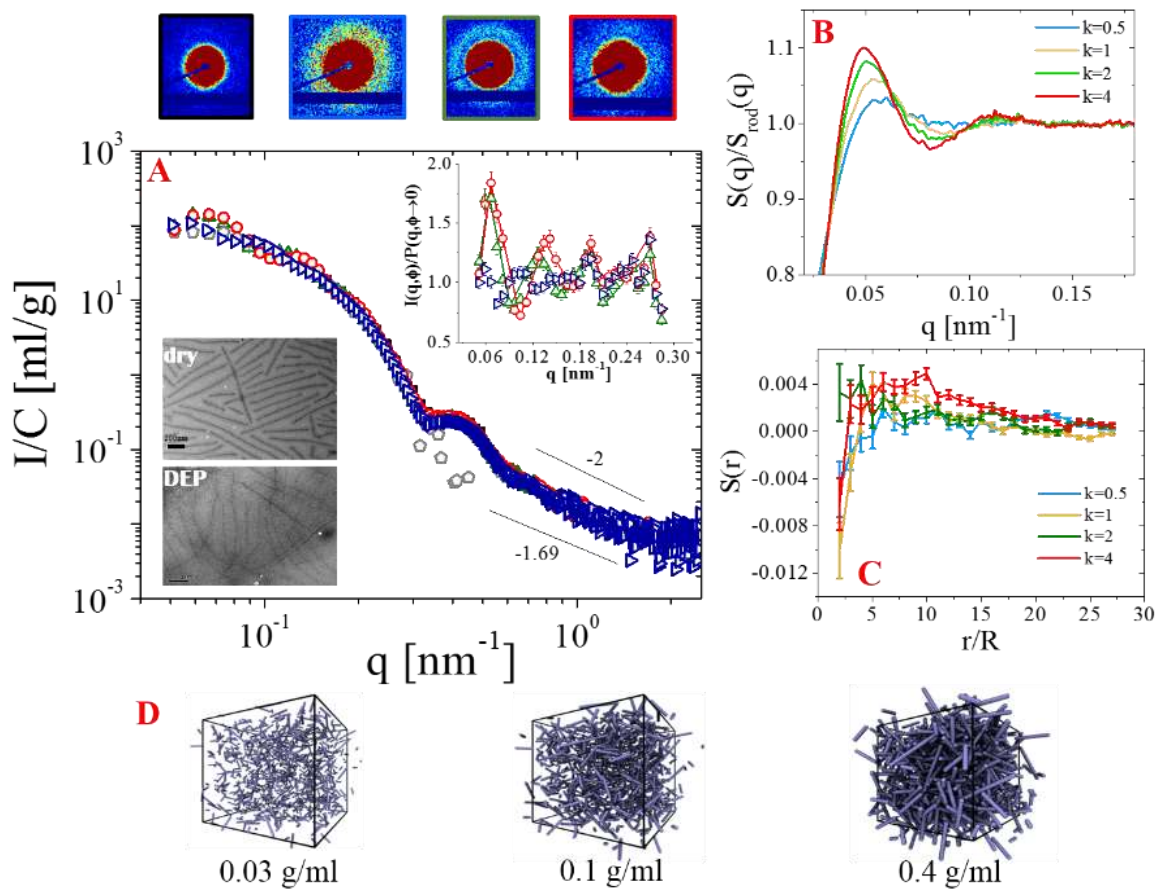

Figure 3. A) SAXS intensity scaled with the nanocylinder concentration against the scattering wave vector at different concentrations in DEP, from dilute to concentrated regime. Grey pentagons, black squares, blue right-pointing triangles, green top-pointing triangles and red circles correspond to $10^{-4}$, $10^{-2}$, 0.1, 0.13, and 0.2 g/ml, respectively. Lines with slopes 1.69 and 2 indicate the two limiting behaviors, self-avoiding and Gaussian (see text). 2D intensity patterns for four selected concentrations are shown on top of the plot from left to right: 0.01, 0.1, 0.13 and 0.2 g/ml. The bottom inset depicts RT- and cryo-TEM micrographs in DEP (0.01 g/ml). The top inset shows the pseudo-structure factor at 0.1 g/ml (blue circles), 0.13 g/ml (green circles), and 0.2 g/ml (red circles), where solid lines are drawn to guide the eye and emphasize the multiple structural peaks. B) Simulated normalized structure factor by that of intra-rod scattering for a suspension of nanocylinders at 0.4 g/ml. C) Simulated orientational correlation function at various values of interparticle potetial's "softness" parameter, k. Error bars represent one standard deviation of the mean over 250 snapshots. D) Snapshots of equilibrated simulation boxes at different concentrations.

## Linear viscoelastic response

The remarkable evolution of oscillatory shear response from liquid to a highly concentrated state is shown in Figure 4. The mechanical spectra cover up to 9 decades in frequency and were obtained by means of a combination of 3 different rheological techniques: high-frequency (above 100 rad/s) piezo-rheometry (see also the SI, section SI8),[52] dynamic frequency sweep and creep measurements (see Figure S7 of the SI) in order to extend the low-frequency region. From 0.03 g/ml to 0.1 g/ml, a dramatic slow-down by 4 decades marks the transition from liquid to dynamically arrested state (whose slow terminal regime is only detected with creep measurements and is assigned to structural relaxation), supporting the picture emerging from Figure 3. This arrested state is attributed to particle crowding, which is well-studied in the context of spherical colloidal glasses and rationalized by invoking the concept of caging[61,64–67]. It has also been observed in fd-viruses above the isotropic-nematic phase coexistence where nematic domains freeze kinetically.[5–7] Kinetic vitrification was also investigated theoretically in isotropic solutions of rigid rods by Yatsenko and Schweizer.[36] They found that the volume fraction at the glass transition decreases as the aspect ratio increases (with L/D>2) and that the shear modulus follows an exponential growth with the volume fraction. Subsequently, Zhang and Schweizer[68] reported that the stress relaxation dynamics of glasses formed by hard rods is increasingly controlled by the center-of-mass translation rather than the rotational motion as the aspect ratio of the rods increases. Moreover, they found that the structural relaxation time grows more steeply than exponentially with the volume fraction. Whereas the present systems bear clear differences, primarily due to softness, there are interesting analogies and similarities with the above theoretical framework.[35,68]

In the present case, in order to rationalize the kinetic arrest of the nanocylinders it is instructive to first invoke the Doi-Edwards theory[69,70] (developed for infinitely thin stiff rods), subsequently extended by Morse for semiflexible rod-like polymers.[71,72] Given that a test cylinder is constrained by the neighbors in a virtual tube, stresses can relax by a fast fluctuation inside the tube and subsequent multiple tube renewal events (long-time diffusion or tube longitudinal motion). By this mechanism one virtual tube renewal involves motion of the test cylinder by a distance equal to its length. Morse[71,72] has shown that the structural relaxation time of a tightly entangled network of semiflexible rods ($\tau_{rod}$) is larger than the time required for a tube renewal event ($\tau_{ren}$). One can interpret $\tau_{rod}$ as the total time within which the test cylinder undergoes multiple tube renewal events in order to completely forget its initial orientation. In the present case, one may adopt this picture of stress relaxation dynamics, of course with the reservation that the above theoretical concepts were developed mainly for long rods.[69–72] Under these conditions, the dynamic response is assumed to be controlled by the confined motion of a single nanocylinder motion and this is due to the fact that the length scale associated with the elasticity of the suspensions is at the level of the single particle radius rather than a domain size, as it will be discussed below. By estimating the number of particles per unit volume as $v=CN_A/M_{cylinder}$ we can infer the concentration regime where a given sample belongs to. Indeed, according to Doi and Edwards,[69] if v is larger than the reciprocal of the pervaded volume that a single rod occupies, $v>1/L^2(2R_c+2h)=1.21 \times 10^{13}$ cm$^{-3}$, the so-called entanglement regime is reached. In our case, such a condition is fulfilled at a concentration of 0.02 g/ml (which actually corresponds to the overlap concentration $C^*$ reported in Figure 1). Note the here there is no liquid crystalline order, as is the case for monodisperse long rods.[69] This picture is suggestive of the highly interactive concentration regime studied and of the constraint dynamics of the crowded nanocylinders. However, the present experimental results are discussed below in the context of decoupled polymeric (shell) and colloidal (nanocylinder) dynamics. Referring to Figure 4, we distinguish different regimes:

**Regime I, dramatic slowing-down of dynamics (0.03-0.1 g/ml):** The two lowest suspensions investigated, 0.03 g/ml and 0.1 g/ml, exceed the abovementioned limit of concentrated isotropic suspensions ($v>1/L^2(2R_c+2h)$), however, no macroscopic order has been observed neither with SAXS nor with simulations. Remarkably, in this concentration regime, the relaxation time grows by four decades, exhibiting a strong concentration dependence that conforms to the predictions of Schweizer and co-workers for hard rods.[36,68] The polydispersity, aspect ratio and softness are at the origin of a more complex particle arrangement where short cylinders form locally favored structures characterized by particle ordering at local scale (as shown in the pseudo-$S(q)$) and domain formation.

**Regime II, evolution of structure (0.1-0.2 g/ml):** From 0.1 to 0.13 g/ml, where a significant difference in birefringence and the onset of structural peaks are observed (see section on structure and phase behavior), the elasticity of the system is strongly affected by the structure as G' increases by one order of magnitude whereas the terminal relaxation times are nearly the same. This can be rationalized by the increase of the parallel translational diffusion coefficient when particles orient locally. Previous studies[73–75] have shown that at the isotropic-nematic transition the translational diffusion speeds-up with concentration as a result of an increase in orientational order parameter and effective free volume due to particle orientation. In the present case though, we postulate that this occurs at local scale due the presence of the multi-domain structure. The elasticity of the suspension (low-frequency colloidal plateau modulus, see below) increases with particle concentration, but the expected slow-down of the structural relaxation is compensated in part by a speed-up of translation due to the local orientation of the nanocylinders.

**Regime III, shell interpenetration, hierarchical dynamics and jamming (0.2-0.6 g/ml):** A further increase in concentration (to 0.2 and 0.23 g/ml) leads to a substantial slow-down of the terminal relaxation time (by two decades), without appreciable change in the elastic modulus. At high enough concentrations, the softness of the system comes into play and shells interpenetrate (see also SI, section SI10 and Figure 2). In fact, shell engagement results in an additional plateau modulus at very high frequencies and an associated stress relaxation due to arm retraction.[76,77] A clear decoupling between colloidal and polymeric dynamics was recently discussed for star polymer melts and star-polymer-linear chain mixtures.[78,79] In the present nanocylinder case, the shell relaxation mode is only present at concentrations of 0.2 g/ml (which is very close to the entanglement concentration of PS polymer chains in athermal solvent,[80] $c_e \approx (N_e/N)^{0.76}=0.24$) or higher. The slow colloidal mode is significantly retarded and conforms to the jamming scenario, which is discussed below (see Figure 5).

By increasing the concentration to 0.4 and 0.6 g/ml, the values of moduli become higher and the terminal dynamics even slower. The hierarchical (polymeric-colloidal) dynamics becomes more evident as two plateau moduli become clearly distinguishable in the mechanical spectra. We attribute the high frequency plateau due to the effective network formed by the strongly engaged shells, hence polymeric in nature, and the low frequency plateau associated with the colloidal response with modulus $G_p$. In fact, once the engaged arms relax, the dynamics is exclusively controlled by the nanocylinder diffusive motion: in-cage fluctuation and tube-renewal dynamics. Note that for these two concentrations the frequency-extend of the colloidal plateau is smaller compared to 0.2 and 0.23 g/ml but the extend of interpenetrated region is larger, due to larger interpenetration which reflects PS entanglements. In addition, a high-frequency crossover was detected as well. We postulate that this very fast characteristic time of the system is purely polymeric in nature and can be attributed to the relaxation of an entanglement strand $(\tau_e)$[80] (see also SI, section SI10).

The value of the colloidal plateau modulus $G_p$ can be used to extract the associated characteristic lengths in the different regimes by considering that a nanoparticle of radius R fluctuates within its effective cage with a fluctuation length:[81] $\xi = \left(\frac{k_B T}{R G_p}\right)^{1/2}$, where $k_B$ is the Boltzmann constant and T the absolute temperature. The obtained values of $\xi$ are 57, 12, 7.2, 6.9, 4.2 and 3 nm for 0.1, 0.13, 0.20, 0.23, 0.4, and 0.6 g/ml, respectively; with the exception of the lowest concentration (with no detectable interpenetration) where $\xi$ is basically the corona radius, the other values of $\xi$ are smaller than the overall corona radius. This suggests that fluctuations of small (local) scale mediate the stress relaxation of the colloidal mode in regimes II and III. Similar approaches have been invoked in the analysis of the relaxation of glassy or jammed soft colloids (stars).[82,83]

**Regime IV, organization in the densely jammed state (0.74 g/ml):** At the highest concentration investigated (0.74 g/ml) the kinetically suppressed multi-domain structure exhibits self-similar viscoelastic response with collapsed moduli exhibiting a 1/3 power-law frequency dependence over four decades in frequency, identical to that of hexagonally packed cylinders from microphase-separated block copolymers.[84] Evidence of a probable hexagonal order has been also found in the X-ray pattern of self-assembled films in the dry state (see Figure S6 of the SI). Nonetheless, we refrain from ascribing this structure to a fully developed hexagonal order due to the presence of a non-negligible polydispersity in length (Table 1), hence we call it hexagonal-like.

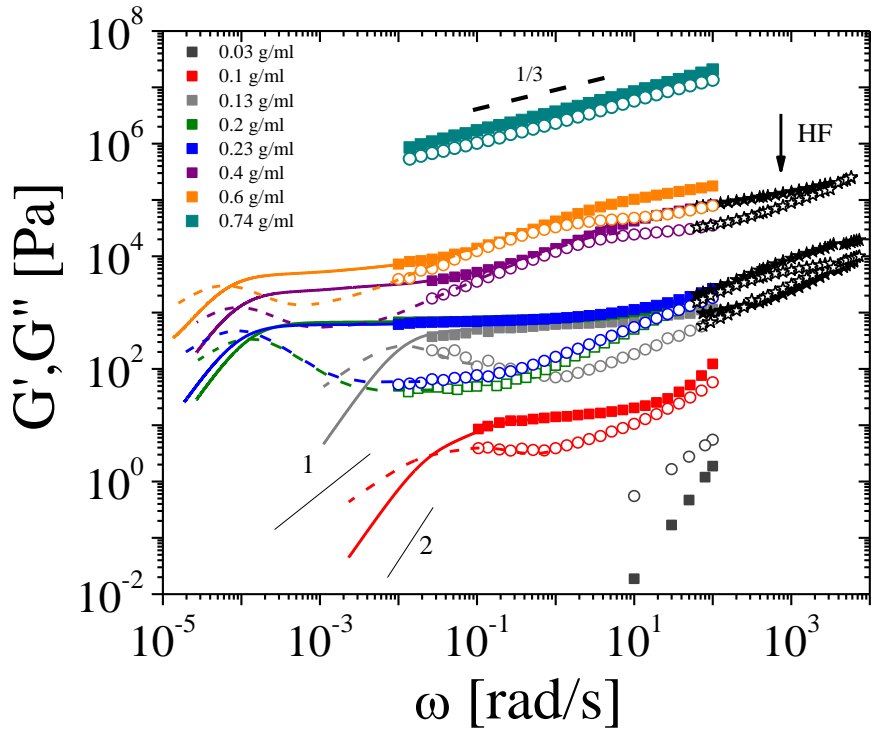

Figure 4. Linear viscoelastic moduli, storage G′ (solid symbols) and loss G″ (open symbols) as functions of the oscillatory frequency at different concentrations. Solid (storage) and dashed (loss) lines represent creep compliance converted into dynamic moduli. High Frequencies above 100 rad/s (HF) were measured with a piezo rheometer at three concentrations (0.13, 0.23 and 0.4 g/ml) and are represented by solid and open stars for G' and G", respectively.[85] The 1/3 slope relates to the expected modulus dependence on frequency for hexagonally packed cylinders in block copolymers.[84]

**Emerging picture:** The structural and the (inside the tube) fluctuation times can, by analogy with the fast in-cage and slow out-of-cage motion for caged spherical colloidal glasses,[64,86] be obtained by the inverse frequency corresponding to the low-frequency moduli crossover ($\tau_{rod}$) and the minimum in G″ ($\tau_\beta$), respectively. Similarly, the colloidal plateau modulus $G_p$ can be estimated as the storage modulus at $\tau_\beta$. These values are reported in Figure 5 as functions of the concentration. At around C=0.1 g/ml, a significant increase of both relaxation times is observed. This coincides with the isotropic to multi-domain structure transition, as also observed with birefringence (Figure 2) and SAXS (Figure 3) measurements. Note that this value coincides with C* if the latter is estimated by considering only the core diameter (27 nm, which is about 1/5 of the corona diameter of 139 nm) for the calculation of the pervaded volume; this value is about five times of the original estimation of 0.02 g/ml, based on the entire diameter. Surprisingly, a further increase in concentration has no strong effect on the terminal time, as well as on the fluctuation time inside the tube. Indeed, both characteristic times tend to level off with increasing concentration. Given their very large values, it is instructive to consider the colloidal plateau modulus, which is also depicted in Figure 5. $G_p$ exhibits a linear dependence on concentration. In fact, the data conform to $G_p=K(C-C_J)$, where K and $C_J$ are fit parameters (see Figure 5). Such a behavior is typical of the so-called jammed state recently observed in microgels[87,88] and even in star-polymer melts.[79] In the present preliminary analysis the $C_J$ is estimated to be 0.11±0.02 g/ml, hence the association of Regime III above with the onset of jamming. Note that jamming in soft hairy materials has been also discussed in the context of activated hoping with theoretical ideas based on mode coupling, and a linear dependence of storage modulus on volume fraction was reported.[89] More generally, it seems that the linear dependence of $G_p$ on concentration is typical of jammed states[87,88,90,91] and the present soft and short cylinders conform to the emerging picture. Such a finding certainly opens a new scenario of jamming in anisotropic colloids at rest, in addition to previous experimental works addressing the jammed state induced by shear flow.[92,93] In this particular situation, we were able to resolve terminal relaxation times in Regime III and note their rather weak concentration dependence, compared to Regimes I, II (Figure 5). This is not unusual for hairy materials with fluctuations of their shell.[94] As a final remark, it is also interesting to remark that if one adapts the Doi-Edwards picture[69,70] of entangled rods, their decaying plateau modulus due to rod orientation scales linearly with concentration (entropic effect). In this context, our short, soft nanocylinders are entangled at C>0.02 g/ml (see also discussion above), however the Doi-Edward theory is not appropriate here as already explained.

To summarize the above observation in the form of a generic state diagram, Figure 6 illustrates a compilation of the viscoelastic properties at different volume fractions (based on Table 2) indicating the different regimes (isotropic – Regime I, isotropic-multi-domain coexistence – Regime II, multi-domain and jammed – Regime III, hexagonal-like order – Regime IV). This confirms the richness of behaviour and associated opportunities for tailoring the properties of such short and soft nanocylinder suspensions.

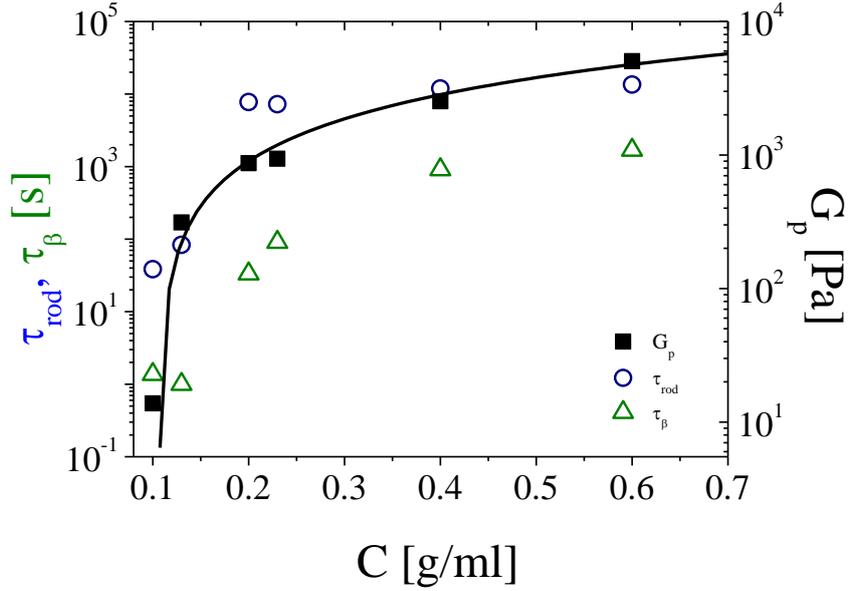

Figure 5. Structural ($\tau_{rod}$) and (inside the tube) fluctuation ($\tau_\beta$) times (left-hand axis) and colloidal plateau modulus $G_p$ (right-hand axis) as functions of concentration. The solid line is the best fit of the plateau modulus to a linear variation within the jammed region. The function has the form[95] $G_p=K(C-C_J)$, where K=9662±727.2 [Pa ml/g] and $C_m$= 0.11±0.02 [g/ml].

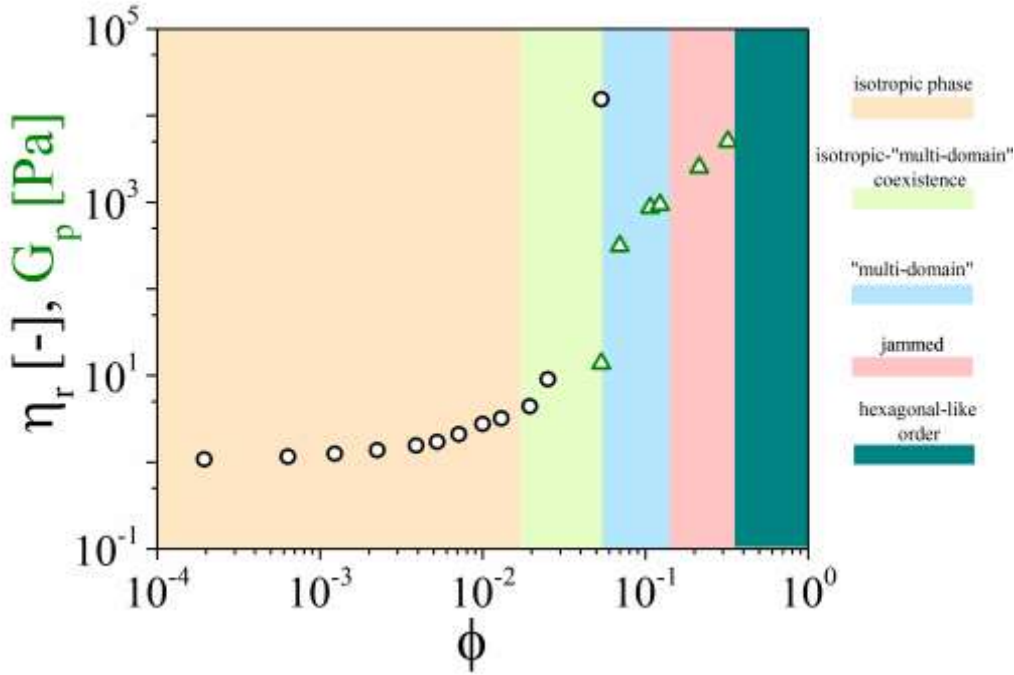

Figure 6: Effective state diagram of the short, soft nanocylinder suspensions (Table 1, Figure 1), reflecting a compilation of rheological data: relative viscosity $\eta_r$ (Figure 2) and colloidal plateau modulus $G_p$ (Figure 4) as functions of the effective volume fraction (Table 2). The regions encompassing different phases or (out-of-equilibrium) states are illustrated with different colors (legend to the right).

# CONCLUSIONS

Whereas rod-like colloidal suspensions exhibit a smooth transition between mesophases upon increasing the volume fraction,[58] the behavior of their short cylindrical counterparts is richer and more complex. Experiments[58] and simulations[8,11] have reported cases where, even for monodisperse cylinders, particles freeze into a macroscopically isotropic state but with local orientational order within domains. When softness is introduced to short cylinders by means of a grafted polymeric shell, their structure and dynamics are greatly affected. In this work, by combining various experimental and computational techniques we have shown that the effect of concentration on the structure of short and soft grafted nanocylinders (with aspect ratio of 6) leads to several dynamic regimes. At the lowest concentrations a dramatic slowing-down of the dynamics is observed; the relaxation time exhibits a strong concentration dependence as predicted for hard rods.[36,68] In the range 0.1 g/mol (about five times the overlap value) to 0.2 g/ml the dynamics follows the structural evolution of the suspension. In this regime, the expected slowing-down of the terminal time is compensated by particle alignment at local scale. At higher concentrations, the suspensions exhibit shell interpenetration and jamming dynamics, with the colloidal plateau modulus depending linearly on concentration, conforming to the emerging picture of jammed (soft) colloidal suspensions. Finally, at 0.74 g/ml, the densely jammed state is characterized by the signature of hexagonally packed cylinders from microphase-separated block copolymers. Above 0.2 g/ml, the contribution of the softness is primarily reflected in shell overlap. The interdigitation of the chains contributes to the hierarchical stress relaxation dynamics involving different length and time scales. Hence, the colloidal dynamics are preceded by arms disentanglement and the associated local motion (clearly detectable at 0.4 g/ml).

The rich dynamics reported in this work certainly reflects the presence of polydispersity in length of the investigated systems, as well as the small aspect ratio and the soft shell (grafted arms), with the latter being associated with the long-ranged repulsive interactions and interdigitation at high concentrations. Isolating experimentally the role of the polydispersity on the dynamics of short core-shell nanocylinders represents an open challenge as it is not trivial to synthesize such systems with a negligible polydispersity in length. Nevertheless, it is hoped that the results presented in this work could provide insights for fundamental understanding in the realm of non-spherical soft colloids as well as opening new routes for tailoring flow properties of suspensions at molecular scale.

## ACKNOWLEDGEMENTS

We thank Thanasis Athanasiou for his help with the high-frequency viscoelastic measurements. This research was partly funded by the EU (European Training Network COLLDENSE (H2020-MCSA-ITN-2014, grant number 642774 and Horizon2020-INFRAIA-2016-1, EUSMI grant no. 731019) and the National Natural Science Foundation of China (Grant No. 21674122). Kevin S. Silmore was supported by the Department of Energy Computational Science Graduate Fellowship program under grant De-FG02-97ER25308.

## SUPPORTING INFORMATION

Analysis of polydispersity, comparison of sizes in different solvents, Dynamic light scattering characterization, estimation of shell thickness, Transmission Electron Microscopy, Small angle X-ray scattering, creep measurements, high-frequency rheometry, Brownian Dynamics Simulations, analysis of shell interpenetration at high concentrations

## REREFENCES